\documentclass[twoside,11pt]{article}
\usepackage{journal2e_v1}

\usepackage{epsf}
\usepackage{epsfig}
\usepackage{amsmath}
\usepackage{subfigure}
\usepackage{amssymb}
\usepackage{algorithm}
\usepackage{multirow}
\usepackage{multicol}
\usepackage{array}
\usepackage{enumitem}
\usepackage{url}

\usepackage{color}
\usepackage[dvipsnames]{xcolor}

\makeatletter
\newcommand{\mylabel}[2]{#2\def\@currentlabel{#2}\label{#1}}
\makeatother

\usepackage{graphicx} 
\usepackage{ltxtable} 
\newcolumntype{L}{>{\centering\arraybackslash} m{0.04\columnwidth}} 
\newcolumntype{R}{>{\centering\arraybackslash} m{0.48\columnwidth}} 
\newcolumntype{S}{>{\centering\arraybackslash} m{0.32\columnwidth}} 

\usepackage{algorithm}
\usepackage{algorithmic}

\usepackage[pagebackref=true,breaklinks=true,letterpaper=true,colorlinks,bookmarks=yes]{hyperref}

\def\0{{\bf 0}}
\def\1{{\bf 1}}

\begin{document}
\title{Methodologies for Improving Modern Industrial Recommender Systems}

\author{\name  Shusen Wang \\
        \addr wssatzju@gmail.com
        }

\date{}

\maketitle

\vspace{-15mm}

\begin{abstract}%
Recommender system (RS) is an established technology with successful applications in social media, e-commerce, entertainment, and more. RSs are indeed key to the success of many popular APPs, such as YouTube, Tik Tok, Xiaohongshu, Bilibili, and others. This paper explores the methodology for improving modern industrial RSs. It is written for experienced RS engineers who are diligently working to improve their key performance indicators, such as retention and duration. The experiences shared in this paper have been tested in some real industrial RSs and are likely to be generalized to other RSs as well.
Most contents in this paper are industry experience without publicly available references.
\end{abstract}


\section{Evaluation Metrics}
\label{sec:eval}

This paper elaborates on the methodologies for improving industrial recommendation systems (RS). 
How does the industry measure whether an experiment makes the RS better or worse?
\begin{itemize} 
\item The most crucial evaluation metrics are traffic and retention. Traffic is measured by daily active users (DAU) and monthly active users (MAU). User retention can be assessed using various metrics; among them, LT7 and LT30 have recently become the most widely accepted retention metrics. In e-commerce, gross merchandise value (GMV) can be as important as the number of users and user retention. 
\item Duration, the number of impressions, and the number of clicks are essential evaluation metrics for RS, although they are less critical than traffic and retention. Duration refers to the total time (typically in minutes) spent on the APP divided by the number of active users. Duration is strongly correlated with retention. The number of impressions (or clicks) largely determines the advertising revenue per user. 
\item User-generated content (UGC) platforms value their authors because the platforms rely on them to create new items to attract users. The number of daily or weekly published items is a vital evaluation metric. On platforms such as YouTube and Tik Tok, ordinary users can also become authors. The participation rate, defined as the daily active authors divided by DAUs, measures the willingness of ordinary users to publish items. 
\item Other metrics, such as click rates, engagement rates, and dislike rates, are monitored. When deciding whether to fully launch an experiment, these metrics are not taken into consideration unless they are negatively affected to a large extent. 
\end{itemize}

It is common for an experiment to increase one metric while decreasing another. If LT grows and DAU is at least neutral, then a decrease in any other metric will be tolerated. However, if both DAU and LT7 are neutral, then the decision to fully launch the experiment will be made on a case-by-case basis.

Despite its popularity, LT may be unfamiliar to many readers. Here, I provide some explanations about LT7; LT30 is defined similarly. Suppose a user logs in to the APP on day $T$. During the 7 days from $T$ to $T+6$, the user logs in to the APP on 4 different days. On day $T$, this user's LT7 is equal to 4. Taking the average over all the active users, we obtain the average LT7 on that day. Naturally, LT7 can range from a minimum of 1 to a maximum of 7.

If users' overall satisfaction improves substantially, then both DAU and LT will increase. However, if users' overall satisfaction remains neutral while satisfaction among inactive users worsens, DAU will decrease, but LT will increase. This occurs because when computing LT, the decrease in the denominator is more significant than the decrease in the numerator. Therefore, an increase in LT alone does not necessarily indicate better user satisfaction; DAU must grow or at least remain neutral.

\section{Retrieval}

Retrieval, also known as candidate generation, is the first step in the RS pipeline. A contemporary RS generally consists of over 20 retrieval models. Two-tower and item-to-item (I2I) models are the most prominent families of retrieval models, collectively accounting for over half of the total quota. While other retrieval models have much smaller quotas, incorporating such models can improve metrics such as retention.

The same retrieval model can be applied to multiple item pools. A retrieval channel is defined as a retrieval model applied to an item pool. A modern RS may contain over 50 retrieval channels. Each retrieval channel has a specific quota; for example, the ItemCF model applied to new items published during the past 7 days has a 5\% quota. It is worth noting that quotas can vary across different user groups.

There are primarily four ways to enhance the retrieval process: improving two-tower models, enhancing I2I models, adding new retrieval models, and incorporating new item pools for specific purposes. In this section, we will discuss the first three methods, while the last one will be revisited in subsequent sections.

\subsection{Two-tower models}

First and foremost, RS engineers should closely examine positive and negative samples, paying special attention to their sources and quotas. The right mixture of training samples is vital for two-tower models.

\begin{itemize} 
\item Simple positive samples consist of user-item pairs with clicks, while simple negative samples are randomly combined user-item pairs. (A user has a very low probability of showing interest in a random item. That is why the random pairs are simple negative samples.) 
\item Hard samples lie between simple positive and simple negative samples. Retrieved items that do not pass pre-ranking can serve as hard negative samples. Utilizing the appropriate hard negative samples is crucial for enhancing two-tower models' performance. The balance between simple negative, hard negative, and positive samples should be carefully adjusted. 
\item A common mistake is using items with impressions but without clicks as negative samples. On the contrary, these can be used as positive samples. An item selected by ranking is likely to align with the user's interests, even if it is not clicked. 
\end{itemize}

There are better neural network architectures than the vanilla two-tower model.
\begin{itemize}
    \item 
    The standard user tower and item tower are simple multi-layer perceptrons. Advanced architectures such as DCN-V2 \citep{wang2021dcn} work better than multi-layer perceptrons.
    \item 
    User’s historically interacted items (which we call last-n) can be the user tower’s inputs. The readers can follow Google’s paper \citep{covington2016deep}. 
    \item 
    The standard user tower outputs a single vector. Its dot product with the item vector estimates how much the user is interested in the item. We call this model single-vector two-tower. Multi-vector two-tower is better in practice. Its user tower outputs multiple vectors, and each vector is responsible for estimating one objective such as click, like, follow, share, comment, etc.
\end{itemize}

Two-tower models can be enhanced with the use of improved loss functions and training methods.
\begin{itemize} 
\item Training a two-tower model using in-batch negative samples has become prevalent since Google's paper \citep{yi2019sampling}. This approach is meant to correct for sampling bias, as popular items are more likely to be both positive and negative samples. Readers can refer to the paper \citep{yi2019sampling} for more details. 
\item Items with a small number of clicks are less likely to appear as positive or negative samples during training, resulting in poorly learned ID embeddings. Google's self-supervised learning method \citep{yao2021self} can address this issue and improve the learning of embeddings for less popular items. 
\end{itemize}

\subsection{Item-to-item (I2I)}

Item-to-item (I2I) is a large family of retrieval models based on item-item similarity. Most I2I models operate as ``user→item→item'', or U2I2I.
Here, ``user→item'' represents an index containing users' interacted items, meaning a user has explicitly expressed their interest in those items. ``item→item'' denotes an index that connects one item to many similar items.
There are various I2I models, and their differences lie in how item-item similarity is measured.
\begin{itemize} 
\item Item-item similarity can be based on user behavior. Two items are considered similar if many users have interacted with both items. 
\begin{itemize} 
\item ItemCF is a classical and widely used method. Swing \citep{yang2020large} is a variant of ItemCF. Both methods are based on the same concept, differing only in the calculation of item-item similarity. 
\item Online ItemCF and online Swing are online learning implementations, as opposed to batch implementations. 
\item All four retrieval models are used in real-world industry RS. They differ enough from one another that an ensemble of these four channels performs better than leaving any of the retrieval channels out. 
\end{itemize} 
\item Item-item similarity can also be based on item embeddings obtained from a two-tower model or graph neural networks. 
\end{itemize}

\subsection{Less popular retrieval models}

An industrial RS typically includes several less important retrieval models, each with quotas less than 5\%. Incorporating a new retrieval model (without increasing the overall number of retrieved items) can improve metrics such as retention. However, adding a new retrieval model inevitably requires more computational resources, such as an additional 1,000 CPU cores.
The following retrieval models have been proven to be helpful.
\begin{itemize} 
\item U2U2I (user→user→item), U2A2I (user→author→item), and U2A2A2I (user→author →author→item) function similarly to U2I2I.
\item 
PDN \citep{li2021path}, developed by Alibaba, jointly models I2I and two-tower.
\item 
Deep Retrieval \citep{gao2021learning}, developed by ByteDance, is deployed in many of its production lines. It is analogous to an earlier work, TDM \citep{zhu2018learning}, which was developed by and deployed within Alibaba. Note that Deep Retrieval and TDM have not been reproduced and validated as useful by any other companies.
\item
Various other innovative retrieval methods such as SINE \citep{tan2021sparse} and M2GRL \citep{wang2020m2grl} are also useful in practice.
\item 
Cache retrieval is not a standard retrieval channel. It caches items that have high-ranking scores but are not displayed due to re-ranking rules. In comparison with other retrieval channels, cached items have a higher likelihood of passing pre-ranking and ranking, and they have a greater chance of being displayed.
\item 
Item cold-start heavily relies on content-based retrieval methods, such as taxonomy, entity, text embedding, image embedding, clustering, etc. This is because newly published items have zero or few interactions with users, making I2I and two-tower methods inapplicable.
\end{itemize}

A real-world industrial RS typically includes tens of retrieval channels. These channels jointly generate a fixed number of items, for example, 5,000 items, which are then sent to preranking. Carefully adjusting the quotas of these retrieval channels can improve key indicators. Additionally, different user groups may have distinct optimal quotas.

\section{Dive Deeper into Deep Learning}

Let's discuss deep learning in more detail. Deep neural networks are applied throughout the RS pipeline. Having these neural networks make more accurate predictions can fundamentally improve the RS.

\subsection{Upgrade ranking models}

In our previous discussion, we covered the two-tower model; thus, in this section, we will focus more on ranking. I will introduce several improvements for ranking models. Once these improvements are implemented, it becomes challenging to further increase metrics such as retention by solely improving ranking models.

Ranking models across the industry exhibit minimal differences; they are fundamentally similar to the wide\&deep model \citep{cheng2016wide}. The inputs include sparse features (e.g., user ID, item ID, taxonomies) and dense features (e.g., user statistics and item statistics). The model comprises multiple heads, each responsible for predicting one target such as click, like, share, follow, comment, etc. Underfitting occurs in ranking models due to the vast amount of training data and the relatively small size of dense layers (which have only a few million parameters). Increasing the depth and width of ranking models usually results in higher accuracy.
The deep component of the wide\&deep model typically has 1 to 6 dense layers, depending on (1) the trade-off between inference cost and prediction accuracy, and (2) the training and inference infrastructure (CPU vs GPU).

Numerous improved neural network architectures have been published. Some of these architectures prove to be effective, while others do not work for unknown reasons.

\begin{itemize} 
\item Automatic feature crossing methods, particularly bilinear crossing \citep{huang2019fibinet}, enhance the accuracy of ranking models without substantially increasing computational costs. Bilinear crossing is effective not only for RS ranking but also for search ranking and ads ranking. LHUC \citep{swietojanski2016learning}, also known as Parameter Personalized Net (PPNet), is another feature-crossing method that improves ranking model accuracy but requires significantly more computation. Parameter and Embedding Personalized Network (PEPNet) \citep{chang2023pepnet} is an upgraded version of PPNet.
\item Ranking models are multitasking. Each of their heads estimates a target such as clicks, likes, shares, follows, comments, etc. A utility function takes these scores as input and outputs a single score that primarily determines the item's ranking. RS engineers continually search for new targets relevant to a user's interests or an item's quality. When they discover a new target, they will add a new head to the ranking model and use the predicted score as a new input to the utility function. 
\item Multi-gate Mixture-of-Experts (MMoE) \citep{ma2018modeling} and Progressive Layered Extraction (PLE) \citep{tang2020progressive} are well-known improvements for multi-task learning. However, it is a common misconception that replacing wide\&deep with MMoE or PLE will immediately result in better prediction accuracy. From my experience and the experiences of many RS engineers, MMoE and PLE do not outperform wide\&deep when holding the computational cost constant. While MMoE and PLE may be helpful in some cases, they are not universally better than wide\&deep. 
\item Position bias is ubiquitous in RS ranking and search ranking. Many papers, e.g., \citep{zhao2019recommending}, have investigated position bias and developed debiasing techniques. I have observed severe position bias in both RS and search contexts. Although many of my colleagues have attempted debiasing, none have succeeded. 
\end{itemize}

Improving metrics such as retention by enhancing neural network architectures is one of the most challenging methods discussed in this paper. Eventually, there comes a point where the potential for further improvement in neural network architecture becomes quite limited.

Modeling users' last-$n$ sequences is a much easier way of improving the ranking model. A user's last-$n$ interacted items are strong indicators of his interests.

\begin{itemize} 
\item We can simply average the $n$ item ID embeddings and use the resulting vector as an input vector for the ranking model. This naive averaging approach works well when $n$ is small. 
\item Deep Interest Network (DIN) \citep{zhou2018deep}, essentially a single-head attention layer, treats the target item (the candidate item to be ranked) as the query and the user's last-$n$ items as keys and values. 
\item Search-based user Interest Modeling (SIM) \citep{pi2020search} is the mainstream approach to long-sequence modeling in the industry. It first uses the taxonomy of the target item to eliminate irrelevant ones from the user's last-$n$ items. It then applies DIN to the remaining $n'$ items ($n' \ll n$). 
\item It is a well-established fact that a larger $n$ leads to more accurate predictions at the cost of increased computation and communication. The challenges of increasing $n$ primarily lie on the infrastructure side rather than the machine learning model side. 
\end{itemize}

The preranking model is a lightweight version of the ranking model. Compared to ranking, preranking deals with 10 times more candidate items, while the per-item computation is 10 times less. The simplest preranking model might be the multi-vector two-tower model, which simultaneously estimates clicks, likes, shares, and all other targets. COLD \citep{wang2020cold}, a three-tower model, is significantly more accurate than the two-tower model; however, it has much higher infrastructure demands.

Maintaining consistency between preranking and ranking is crucial for the overall performance. If preranking and ranking were significantly different, items that should rank high might be eliminated by preranking.
\begin{itemize} 
\item During training, instead of fitting the ground truth $y$, we allow the preranking model to fit $\frac{y+p}{2}$, where $p$ is the prediction made by the ranking model. For example, if $y=1$ means the user actually clicked the item and the predicted CTR is $p=0.6$, we use $\frac{y+p}{2} = 0.8$ instead of $y=1$ as the training label. 
\item Listwise training of the preranking model is also prevalent in the industry. Given a list of $k$ candidate items, we sort the items according to the predictions made by the ranking model. Then, we let the preranking model fit the order of the $k$ items. This approach is similar to the ``learning to rank'' of search engines. 
\item Although the aforementioned methods improve metrics such as retention, they make the system less reliable by propagating ranking model mistakes to preranking. When the ranking model goes wrong, perhaps due to infrastructure failures, the preranking training data may be contaminated. 
\end{itemize}

\subsection{Online learning}

Ranking, preranking, and two-tower retrieval models need to be updated at least once per day. For example, at midnight, the impressions, clicks, and engagements that occurred during the past 24 hours are used to create training samples. Then, the model is incrementally trained for one epoch based on the previous day's checkpoint.

Online learning updates the model with a much higher frequency. There are two major benefits to online learning. First, a user's most recent interests are captured not only by their last $n$ items but also by their user ID embedding, resulting in recommendations that better align with the user's latest interests. Second, newly published items' ID embeddings are initialized with a default vector, which can lead to inaccurate predictions of clicks and engagements. Updating a new item's ID embedding soon after clicks and engagements occur will make subsequent recommendations more accurate.

Despite its benefits, online learning has several shortcomings.

\begin{itemize} 
\item One drawback is the susceptibility of online learning to infrastructure failures. This vulnerability arises because online learning relies on a streaming data pipeline and requires frequent training and deployment of models. 
\item A more severe shortcoming is that online learning can significantly slow down model development. For instance, if the RS comprises one holdout model, one baseline model, and four groups of experimental models, six sets of expensive CPU/GPU clusters would be required to train the six models online. Note taht even with six sets of CPU/GPU clusters, only four different models can be experimented with simultaneously. 
\end{itemize}

Implementing online learning in an immature RS can be detrimental. It can substantially impede model development and introduce numerous stability issues. In summary, consider deploying online learning only when your models have reached a high level of performance and become hard to further improve.

\section{Increase Diversity}

Enhancing the diversity of the displayed items will lead to improved user satisfaction. I am confident that seasoned recommendation system engineers know how to encourage diversity in the final ranking. In fact, there are methods to boost diversity throughout the entire pipeline, including retrieval, pre-ranking, and ranking stages.

\subsection{Diversity in retrieval}

The two-tower and U2I2I retrieval models mentioned above are deterministic. However, in practice, introducing some randomness can enhance item diversity, consequently improving metrics such as retention.

First, let's discuss the two-tower retrieval model. During the retrieval process, the user tower takes the user's features as input and generates a vector representation of the user.
\begin{itemize} 
\item The user vector is utilized to retrieve items from a vector database. Surprisingly, injecting random Gaussian noise into the vector can actually improve the key performance indicators of the recommendation system! To be more specific, if a user has a narrow spectrum of interests, for instance, the user's last $n$ items cover only a small number of taxonomies, then the random noise should be stronger. 
\item The user tower takes the user's last $n$ interacted items as input. Instead of using all the items, we retain the most recent $r$ ($\ll n$) items and randomly sample a subset from the remaining $n-r$ items. This method introduces additional randomness into the user's vector representation. 
\end{itemize}

For the U2I2I retrieval model, we should focus on diversifying the former ``I'', which refers to the user's last $n$ interacted items. The last $n$ items may be concentrated within a small number of taxonomies, for example, 90\% of the $ n$ items belonging to just 3 taxonomies. This can result in a lack of diversity among the retrieved items.

To address this issue, we can set a limit on the size of each taxonomy and down-sample items accordingly. For instance, we could limit the size to 10 items. Among the last $n$ items, let's say 30 items belong to the ``movie'' taxonomy and 6 belong to the ``sport'' taxonomy. In this case, we would randomly sample 10 items from the ``movie'' category and discard the remaining 20, while retaining all 6 ``sport'' items.

\subsection{Diversity in ranking}

Ranking and pre-ranking are based primarily on predicted click and engagement probabilities, and secondarily on diversity scores and rules. In this discussion, we will explore two approaches to diversity within ranking and pre-ranking.

\begin{itemize} 
\item Maximal Marginal Relevance (MMR) \citep{carbonell1998use} and Determinantal Point Process (DPP) \citep{chen2018fast} are the most popular methods for increasing diversity. Let $i$ be an item, $d_i$ the diversity score computed by MMR or DPP, and $s_i$ the score output by the utility function (by combining click and engagement probabilities). We use either $s_i$ or $d_i + s_i$ to rank candidate items. Ranking and pre-ranking present some subtle differences: 
\begin{itemize} 
\item In ranking, we always use $d_i + s_i$ to rank candidate items. In pre-ranking, we initially use $s_i$ to rank candidate items and select the top items that best match the user's interests. We then utilize $d_i + s_i$ to rank the remaining items and select the top ones. 
\item Sliding window is an optional setting for MMR and DPP. It is used in ranking but not in pre-ranking. 
\end{itemize} 
\item MMR and DPP are also known as ``soft scattering''. In ranking, we require ``hard scattering'' to ensure that neighboring displayed items are not overly similar. For instance, items from the same taxonomy must be separated by at least five items. In addition to taxonomy, we also use item clustering to scatter items. Specifically, we can use the embedding of text, images, and videos to partition all the items into 1000 clusters, performed offline. During ranking, items in the same cluster must be separated by at least five items. 
\end{itemize}

In addition to the aforementioned diversity methods, modern recommendation systems (RS) may reserve 2\% of impressions for interest exploration. For example, we maintain a high-quality item pool for male users aged 30 to 40 years, based on item content quality, author quality, and engagement statistics calculated on the same user group. When making recommendations for such a user, 2\% of the impressed items are selected from this item pool.

Since it is weakly personalized, interest exploration may slightly decrease retention in the short term. However, over the long term, retention will gradually recover and ultimately surpass the control group.

\section{Take Care of Special User Groups}

New and inactive users require special attention for two reasons. Firstly, these users have a higher likelihood of churning. For them, the primary goal of the RS is retention, while duration, reading, purchasing, and ad clicks are not considered at all. Secondly, the behavior of these users differs significantly from that of long-term users. On one hand, ranking models trained on data from all users tend to perform poorly when applied to new and inactive users. On the other hand, specific strategies can be employed to engage and retain these users.

\subsection{Specialized item pools}

New and inactive users often have a limited number of clicks and engagements, which can result in less accurate personalized recommendations. To improve their satisfaction, the focus should be on recommending high-quality items that may or may not match their specific interests.

Users can be divided into groups based on demographic data. For instance, males between the ages of 30 and 40 can form a user group, and a pool containing items likely to match their interests can be created. The simplest method for selecting high-quality items is to use engagement statistics such as like rate, follow rate, share rate, and comment rate. A more advanced technique is causal inference, which identifies content or statistical features that contribute to user retention.

Although there may be several item pools, the same two-tower model can be applied across all of them. It is not necessary to train a separate two-tower model for each individual item pool. During the inference process, the item tower computes the item embeddings, and an Approximate Nearest Neighbor (ANN) index is built for each item pool. Since each item pool is smaller than the main pool by orders of magnitude, constructing an ANN index for each individual item pool is relatively inexpensive.

\subsection{Specialized ranking strategies}

There are several protective strategies for enhancing the satisfaction of new and inactive users. The quantity of advertisements and other low-quality items should be strictly controlled or even reduced to zero. Profiting from these users is totally unnecessary; instead, the focus should be on preventing them from churning.

Newly published items often lack user interaction and may not be recommended to the right audience. Boosting new items during the ranking process increases their chances of being viewed. After several trials, the RS can identify which users are interested in such new items. However, since initial recommendations may be inaccurate, it is best to avoid testing them on new and inactive users. For these users, new items should not be boosted during the ranking process, ensuring that they are treated equally as older items.

A significant portion of new and inactive users may not even be inclined to click on any item, let alone like, share, comment, or follow. Encouraging these users to click is essential, as disinterest may lead to churning. Consequently, in the ranking utility function, the weight of a click should be set higher for new and inactive users in comparison to long-term users.

\subsection{Specialized ranking models}

Special users exhibit behavior that is quite distinct from that of the majority of users. For instance, their click rates (\#clicks/\#impressions) may be lower than those of typical users, while their like rates (\#likes/\#clicks) tend to be higher. When a ranking model is trained on data from all users, it may be dominated by the behavior of active users, leading to an overestimation of click rates and an underestimation of like rates for special users. Several methods can be employed to calibrate the ranking model's predictions.

\begin{itemize} 
\item Two ranking models can be used: a large model for all users and a smaller model for special users. Both models are DNNs with large-scale sparse embeddings, but they differ significantly in size. 
\begin{itemize} 
\item The large model is trained on data from all users. Predictions for active users are made using only this model. 
\item A smaller model for special users is trained to fit residuals. For a special user, let $y$ be the true label and $p$ be the prediction made by the large model. The smaller model is trained to fit the difference $y-p$. When ranking for a special user, the predictions from both models are combined. 
\end{itemize} 
\item Alternatively, a Gradient Boosting Decision Tree (GBDT) can be used for calibration. The large model predicts click and engagement rates. The smaller model then takes the predicted rates and the user's features as input and outputs calibrated predicted rates. 
\item A single large-scale DNN with multiple experts (similar to MMoE) can be used. The weight of the experts depends on the user's features. 
\end{itemize}

A common pitfall is assigning an individual ranking model to each user group. In the short term, this approach can produce more accurate predictions, thereby improving key metrics such as retention. However, maintaining multiple ranking models increases the complexity of updating the RS. After adding a feature or updating neural network architecture, an A/B test must be performed for each model. If RS engineers continuously update the main model while neglecting the special user groups' models, those models will eventually become outdated, hindering their performance compared to the main model.

\section{User Engagements}

Three crucial engagement activities---following, sharing, and commenting---are goldmines in RSs. In advanced industrial RSs, the DNN models can hardly be improved without incurring significant resource costs. In very recent years, the most advanced industrial RSs are primarily improved by leveraging these three engagement activities rather than updating the DNN models.

\subsection{Follow}

In social media RSs, a user can opt to follow an author if they find the content interesting. 
For a new user, the number of authors they follow (denoted by $f$) has a positive correlation with their retention rate (denoted by $r$). The relationship between $f$ and $r$ is particularly strong when $f$ is low. When $f$ reaches a large enough value, increasing $f$ has minimal impact on $r$. Several methods can enhance new users' retention by exploiting the following engagement, and we will discuss two below:
\begin{itemize}
\item
Add an additional term, $w(f) * p$, to the utility function. Here, $w(f)$ represents the weight that decreases as $f$ increases, and $p$ denotes the predicted probability that the user will follow the author. If a new user does not follow many authors, the RS tries to encourage further following engagement.
\item
Maintain a pool of items with high follow rates (\#follow/\#click and \#follow/\#impression). For new users, a selection of items is retrieved from this pool. Upon viewing the items, users might follow the authors. As they follow more authors, the likelihood of user churn decreases.
\end{itemize}

Modeling the following relationship is beneficial not only for new users but for all users in general. Typically, a user is more likely to click on items published by authors they follow.
\begin{itemize}
    \item 
    The following relationship can be used in retrieval, e.g., user → author → item and user → author → author → item.
    \item 
    Sometimes, user U is highly likely to click items published by author A, but for some unknown reasons, U does not click the follow button on A's profile. We refer to this as implicit following. The number of implicitly followed authors can be many times larger than the explicitly followed authors. The implicit following relationship is very useful in retrieval.
\end{itemize}

User-generated content (UGC) platforms, such as TikTok, encourage authors to publish items. The number of daily published items is one of the key evaluation metric of a UGC RS. The following relationship can be utilized to motivate authors to publish more content. In particular, when an author has few followers, an increase in his follower count can strongly encourage him to publish. RSs should assist these authors in attracting more followers. For instance, if an item is published by such an author, the predicted following rate should be given a higher weight.

\subsection{Share}

A YouTube user may share a video on Twitter, which generates new traffic or even new user registrations for YouTube.\footnote{Disclaimer: I use YouTube as an example. I am not aware of any mechanism employed by YouTube specifically to encourage users to share videos on other platforms.} Several social media RSs in China are actively working to promote sharing to increase their DAU.

By increasing the weight of the predicted sharing rate (\#share/\#click) in the utility function, the number of shares will naturally increase. The decision to boost the weight of the predicted sharing rate depends on whether the user is a key opinion leader (KOL).
\begin{itemize}
    \item 
    If the user is a KOL, for instance, with 100K followers on Twitter, his sharing of a YouTube video on Twitter could bring a large amount of traffic to YouTube. If the RS knows the user is a KOL, it should exploit the user's value by boosting the weight of sharing.
    \item 
    A user might have 100K followers on Twitter but zero followers on YouTube. How does YouTube identify a KOL? The answer is simple: if the user's past shares have generated significant traffic, he is likely a KOL. The estimated KOL score should be recorded in the user's profile.
\end{itemize}

Additionally, a pool of items likely to be shared on external social media platforms can be established. When recommending content to KOLs, extra items are retrieved from this pool to encourage sharing.

\subsection{Comment}




On social media platforms, users often leave comments on items they like or dislike. Recent industry research (which will not be published) indicates that comments have significant value in RS. Here, I discuss some preliminary findings.

Comments can significantly encourage authors to publish more item. For instance, suppose you create a video and upload it to YouTube. If your video quickly receives several positive comments, you will be motivated to produce more videos. Indeed, following and commenting are the most crucial engagement activities that encourage authors to publish. In ranking, if an item has been engaged (e.g., liked and shared) but not commented on, the weight of the predicted comment rate should be boosted in the utility function.

Users who frequently leave comments are less likely to churn. Specifically, some users have ample leisure time and are willing to leave comments. RSs should allow these users more opportunities to comment and participate in discussions. To achieve this, an item pool with a high number and rate of comments can be maintained. These users can access a customized retrieval channel based on the item pool, providing more opportunities for interaction.

\section{Concluding Remarks}

This paper summarizes my understanding of how to improve an industrial Recommendation System (RS). At the early stage of developing an RS, efforts should be focused on improving models. Without reliable predictions of clicks and engagements, many of the methods introduced in this paper will not function effectively. Once the two-tower and I2I retrieval models, the three-tower pre-ranking model, and the wide\&deep multi-task ranking model are established, the RS can be considered modern.

After numerous updates, improving the pre-ranking and ranking models through feature engineering and enhanced neural network architectures becomes increasingly difficult. At this point, it is time to consider online learning and long-sequence modeling (e.g., SIM); these approaches will significantly increase the accuracy of the models' predictions. Whether they will be implemented depends on the strength of your infrastructure and how you balance costs and returns.

Eventually, particularly after the launch of online learning, it becomes incredibly challenging to further enhance the ranking and pre-ranking models. In this case, you may need to rely on incorporating more advanced retrieval models and item pools, enhancing diversity, implementing special treatment for specific user groups, and fully utilizing user engagements.

\bibliography{bib/rs}

\end{document}